\documentclass[10pt,conference]{IEEEtran}
\IEEEoverridecommandlockouts{}

\PassOptionsToPackage{dvipsnames}{xcolor}
\usepackage{colortbl}
\usepackage{graphicx}
\usepackage{color}
\usepackage{ifthen}
\usepackage{multirow}
\usepackage[inline]{enumitem}
\usepackage[ruled,linesnumbered]{algorithm2e}
\usepackage{float}
\usepackage{orcidlink}

\usepackage{tcolorbox}
\tcbuselibrary{breakable}

\usepackage{booktabs}
\usepackage{tabularx}
\newcolumntype{L}{>{\raggedright\arraybackslash}X}
\newcolumntype{R}{>{\raggedleft\arraybackslash}X}
\newcolumntype{C}{>{\centering\arraybackslash}X}

\usepackage{listings}
\usepackage{hyperref}
\hypersetup{
	breaklinks=true,
	colorlinks=true,
	linkcolor=violet,
	citecolor=violet,
	urlcolor=violet,
}

\usepackage{relsize}
\usepackage{threeparttable}
\usepackage{amsmath}
\usepackage{mathtools}
\usepackage{fontawesome}
\usepackage{xspace}
\usepackage{pbalance}

\usepackage[capitalise]{cleveref}
\usepackage{cite}

\makeatletter
\renewcommand{\paragraph}[1]{\textbf{#1.}\space}
\makeatother

\usepackage[T1]{fontenc}
\usepackage{inconsolata}

\newcommand{\eg}[1]{(e.g., #1)}
\newcommand{\ie}[1]{(i.e., #1)}
\newcommand{\etal}[0]{et al.}

\newcommand{\lspfuzz}{\textsc{LspFuzz}\@\xspace}
\newcommand{\lspfuzznc}{\textsc{LspFuzz}\textsubscript{NC}\@\xspace}
\newcommand{\artifact}{\url{https://doi.org/10.5281/zenodo.17052142}}

\newcommand{\reportedBugs}{51}
\newcommand{\confirmedBugs}{42}
\newcommand{\fixedBugs}{26}
\newcommand{\cves}{two}

\newcommand{\locLspfuzz}{12,293}

\newcommand*{\binaryBaseline}{\textsc{LspBin}}
\newcommand*{\grammarBaseline}{\textsc{LspGram}}
\newcommand*{\twoDimBaseline}{\textsc{Lsp2D}}

\title{\lspfuzz{}: Hunting Bugs in Language Servers}
\author{
	\IEEEauthorblockN{
		Hengcheng Zhu\orcidlink{0000-0002-3082-5957}\IEEEauthorrefmark{1}, Songqiang Chen\orcidlink{0000-0002-1220-8728}\IEEEauthorrefmark{1}, Valerio Terragni\orcidlink{0000-0001-5885-9297}\IEEEauthorrefmark{2}, Lili Wei\orcidlink{0000-0002-2428-4111}\IEEEauthorrefmark{3} \\ Yepang Liu\orcidlink{0000-0001-8147-8126}\IEEEauthorrefmark{4}, Jiarong Wu\orcidlink{0000-0001-6126-303X}\IEEEauthorrefmark{1}, and Shing-Chi Cheung\orcidlink{0000-0002-3508-7172}\IEEEauthorrefmark{1}
		\thanks{\IEEEauthorrefmark{1}%
			Shing-Chi Cheung is the corresponding author of this paper.
			\IEEEauthorrefmark{4}%
			Yepang Liu is affiliated with both the Research Institute of Trustworthy Autonomous Systems and the Department of Computer Science and Engineering.
		}
	}
	\IEEEauthorblockA{
		\IEEEauthorrefmark{1}The Hong Kong University of Science and Technology, Hong Kong SAR\\
		\IEEEauthorrefmark{2}University of Auckland, Auckland, New Zealand,
		\IEEEauthorrefmark{3}McGill University, Montreal, Canada\\
		\IEEEauthorrefmark{4}Southern University of Science and Technology, Shenzhen, China
	}
	Emails: \IEEEauthorrefmark{1}\{hzhuaq, i9s.chen\}@connect.ust.hk, \{jwubf, scc\}@cse.ust.hk\\
	\IEEEauthorrefmark{2}v.terragni@auckland.ac.nz, \IEEEauthorrefmark{3}liil.wei@mcgill.ca, \IEEEauthorrefmark{4}liuyp1@sustech.edu.cn
}

\def\showcomments{}

\ifx\showcomments\undefined
	\newcommand{\editnote}[3]{}
	
\else
	\newcommand{\editnote}[3]{\xspace\colorbox{#1}{\sffamily \smaller \textcolor{white}{~\faCommenting{}~#2~}}\textcolor{#1}{~{#3}}\xspace}
	
\fi

\definecolor{nord0}{HTML}{2E3440}
\definecolor{nord1}{HTML}{3B4252}
\definecolor{nord2}{HTML}{434C5E}
\definecolor{nord3}{HTML}{4C566A}
\definecolor{nord4}{HTML}{D8DEE9}
\definecolor{nord5}{HTML}{E5E9F0}
\definecolor{nord6}{HTML}{ECEFF4}
\definecolor{nord7}{HTML}{8FBCBB}
\definecolor{nord8}{HTML}{88C0D0}
\definecolor{nord9}{HTML}{81A1C1}
\definecolor{nord10}{HTML}{5E81AC}
\definecolor{nord11}{HTML}{BF616A}
\definecolor{nord12}{HTML}{D08770}
\definecolor{nord13}{HTML}{EBCB8B}
\definecolor{nord14}{HTML}{A3BE8C}
\definecolor{nord15}{HTML}{B48EAD}

\definecolor{duck}{HTML}{D89D00}

\newcommand{\code}[1]{{\small\color{violet}\ttfamily{}#1}}

\definecolor{summarybg}{HTML}{E1EDFC}
\definecolor{findingbg}{HTML}{FFF8E6}

\newcounter{rq}
\newenvironment{summary}{
	\begin{tcolorbox}[
			arc=2mm,
			boxrule=0.0pt,
			left=2pt,
			right=2pt,
			top=2pt,
			bottom=2pt,
			colback=summarybg,
			colframe=summarybg,
			enlarge top by=-0.3em,
		]
		\textbf{\faLightbulbO\ RQ\refstepcounter{rq}\therq{} Takeaway:}
		}{\end{tcolorbox}}

\definecolor{changesbg}{HTML}{FFF8E6}
\definecolor{changestitle}{HTML}{FF894F}

\definecolor{commentsbg}{HTML}{DDF4E7}

\begin{document}
\maketitle

\begin{abstract}
	The Language Server Protocol (LSP) has revolutionized the integration of code intelligence in modern software development.
There are approximately 300 LSP server implementations for various languages and 50 editors offering LSP integration.
However, the reliability of LSP servers is a growing concern, as crashes can disable all code intelligence features and significantly impact productivity, while vulnerabilities can put developers at risk even when editing untrusted source code.
Despite the widespread adoption of LSP, no existing techniques specifically target LSP server testing.
To bridge this gap, we present \lspfuzz{}, a grey-box hybrid fuzzer for systematic LSP server testing.
Our key insight is that effective LSP server testing requires holistic mutation of source code and editor operations, as bugs often manifest from their combinations.
To satisfy the sophisticated constraints of LSP and effectively explore the input space, we employ a two-stage mutation pipeline: syntax-aware mutations to source code, followed by context-aware dispatching of editor operations.
We evaluated \lspfuzz{} on four widely used LSP servers.
\lspfuzz{} demonstrated superior performance compared to baseline fuzzers, and uncovered previously unknown bugs in real-world LSP servers.
Of the \reportedBugs{} bugs we reported, \confirmedBugs{} have been confirmed, \fixedBugs{} have been fixed by developers, and \cves{} have been assigned CVE numbers.
Our work advances the quality assurance of LSP servers, providing both a practical tool and foundational insights for future research in this domain.

\end{abstract}

\begin{IEEEkeywords}
	Language Server Protocol, Fuzzing, Software Testing, Developer Tools
\end{IEEEkeywords}

\section{Introduction}\label{sec:introduction}

Modern software development increasingly relies on code intelligence features such as real-time error detection, auto-completion, and refactoring.
The Language Server Protocol (LSP)~\cite{link:lsp} has revolutionized the delivery of these capabilities.
At its core, LSP defines a standardized communication protocol in which code editors act as \emph{LSP clients} and language-specific tools serve as \emph{LSP servers}~\cite{book:lsp-impl}, fostering a robust and interoperable ecosystem of development tools.
LSP is widely adopted, with approximately 300 LSP server implementations supporting a wide range of programming languages and around 50 editors \eg{\textsc{VSCode}, \textsc{Neovim}, \textsc{Cursor}, \textsc{Zed}} offering LSP support.
Given this widespread adoption, an increasing number of developers have incorporated LSP into their workflows.
For example, LSP-related \textsc{VSCode} extensions have attracted millions of installs~\cite{link:vscode-extensions}.

\begin{figure}
	\centering
	\includegraphics[width=0.9\linewidth]{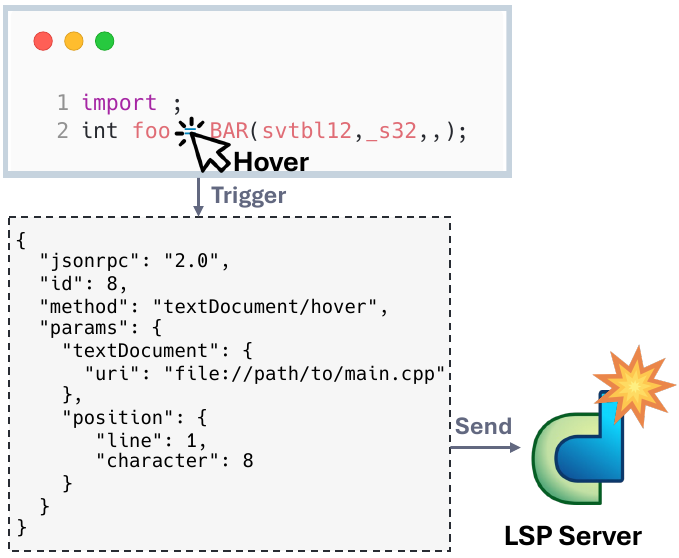}
	\caption{Hovering over the Equal Sign Causes a Crash in \texttt{clangd} 20.1.4}
	\label{fig:clangd-crash}
\end{figure}

However, the reliability of LSP servers is a growing concern.
Crashes in LSP servers can strip off all code intelligence features, leaving developers unproductive and facing unexpected disruptions.
For instance, \cref{fig:clangd-crash} illustrates a crash in \texttt{clangd} (a C/C++ LSP server by LLVM)~\cite{link:llvm-138096}.
When developers write a variable initialization statement and hover the mouse cursor over the equal sign to view information about the assignment operator, \texttt{clangd} unexpectedly crashes.
As a result, developers not only fail to obtain the desired information but are also confused by the sudden loss of all code intelligence features.
Bugs in LSP servers are prevalent, with \texttt{clangd} alone having over 500 issues reporting crashes.
Developers express frustration at the disruption caused by such crashes in LSP servers, with some complaining that it \emph{crashed 5 times in the last 5 minutes}\footnote{\url{https://www.reddit.com/r/neovim/comments/1hxcs1q}}, while others report the exasperating experience of \emph{losing all its parsing and jump-to-definition functionalities}\footnote{\url{https://github.com/clangd/clangd/issues/2070}}.
Moreover, we found that some memory corruptions in LSP servers may lead to security vulnerabilities, such as remote code execution when processing malicious source files.

Despite the importance of assuring LSP server quality, to the best of our knowledge, no existing techniques specifically target LSP server testing.
This motivated us to bridge this gap.

Developing an effective LSP server testing technique requires tackling two unique challenges.
First, we must be able to generate test cases that satisfy the combinatorial input constraints \eg{in \cref{fig:clangd-crash}, the \code{position} should point to a valid location in the source file}.
Without systematic awareness of these constraints, existing binary~\cite{DBLP:conf/woot/MaierEFH20} and grammar-based~\cite{DBLP:conf/ndss/AschermannFHJST19,DBLP:conf/issta/SrivastavaP21,DBLP:conf/sigsoft/SteinhofelZ22,DBLP:journals/pacmse/AmayaSZ25} fuzzers struggle to penetrate beyond input validation layers to exercise the core functionality of LSP servers.
Second, effective testing requires exploring diverse combinations across two key dimensions:
\begin{enumerate*}
	\item the source code being analyzed by the LSP server, and
	\item the editor operations performed on that code.
\end{enumerate*}
The crash in \cref{fig:clangd-crash} exemplifies this challenge: the bug only manifests through the interaction between specific code content \ie{an invalid function call} and a particular editor operation \ie{hovering over the equal sign}.
This multiplicative relationship between source code and editor operations demands a testing approach that systematically explores their interaction to enhance bug-finding capability.

To tackle these challenges and test LSP servers effectively, we propose \lspfuzz{}, a grey-box hybrid fuzzer designed for LSP servers.
Our key insight is that \emph{effective LSP server testing requires the holistic mutation of source code and editor operations}, as bugs often manifest from their specific combinations.
In \lspfuzz{}, our mutation operators are aware of the sophisticated constraints imposed by the LSP, which enables us to produce test cases that can penetrate the input validation layer and reach the core logic of LSP servers.
At the core of \lspfuzz{} is a two-stage mutation pipeline that performs context-aware mutation of the source code and editor operations.
Specifically, the first stage performs syntax-aware mutations to the source code with the \textsc{Tree-Sitter}~\cite{software:tree-sitter} grammar, producing diverse source code as a basis.
The second stage dispatches editor operations based on the mutated source code to trigger various LSP server behaviors.
To achieve this, we leverage syntactic characteristics and LSP server responses to identify locations that can lead to distinct and deeper LSP server behaviors.
With these strategies, \lspfuzz{} can effectively explore the combinatorial space of source code and editor operations to uncover bugs in LSP servers.

In our evaluation, we ran \lspfuzz{} on four widely used LSP servers, such as \texttt{clangd} in LLVM and \texttt{sorbet} developed by Stripe.
The experimental results demonstrate that \lspfuzz{} can effectively detect bugs in LSP servers.
On average, \lspfuzz{} detected 54.1 crashes at different crash locations in these LSP servers across 10 repetitions.
Moreover, \lspfuzz{} achieves higher code coverage compared to baseline fuzzing tools, with improvements ranging from 2.45x to 142.9x over baseline approaches.
Our analysis shows that the two-stage mutation pipeline is essential for effectiveness, especially for LSP servers with rich editor operation support.
We reported \reportedBugs{} bugs to the respective development teams and received positive feedback.
At the time of writing, \confirmedBugs{} have been confirmed, \fixedBugs{} have been fixed, and \cves{} have been assigned CVE numbers.
In addition, LLVM developers proactively responded to our security advisories by disabling \texttt{clangd} in their VSCode extensions for untrusted workspaces (\cref{sec:rq4}).
A developer also expressed interest in integrating \lspfuzz{} into their testing workflows.

In summary, this paper makes the following contributions:
\begin{itemize}
	\item To the best of our knowledge, we are the first to systematically explore the quality assurance of LSP servers.
	\item We designed and implemented \lspfuzz{}, a novel hybrid fuzzer for LSP servers, with \locLspfuzz{} lines of Rust code.
	\item We evaluated \lspfuzz{} on four widely used LSP servers, showing that it significantly outperforms baselines, and that the two-stage mutation pipeline is essential.
	\item We reported \reportedBugs{} previously unknown bugs to the vendors of LSP servers, with \confirmedBugs{} confirmed, \fixedBugs{} fixed, and \cves{}
	      CVEs granted.
	      We received positive feedback from LSP server developers.
	\item We made \lspfuzz{} and our experimental data public~\cite{artifact:zenodo} to facilitate future research.
\end{itemize}

\section{Background}\label{sec:background}

The Language Server Protocol (LSP)~\cite{link:lsp}, introduced by Microsoft in 2016, provides a standardized interface between code editors and language analysis tools.
LSP decouples code editors from language-specific analysis logic, allowing any editor to support languages with LSP servers and any language with an LSP server to provide code intelligence features across all LSP-compatible editors.
This architecture has made LSP the de facto standard for code intelligence features~\cite{book:lsp-impl}.
Around 50 code editors \eg{\textsc{VSCode}, \textsc{Neovim}, \textsc{Zed}} include LSP clients, and there are over 300 LSP servers for various languages, such as \texttt{clangd} for C/C++ and \texttt{sorbet} for Ruby.
This widespread adoption establishes LSP as the standard protocol for editor-language integration.

LSP servers work with code editors in a dynamic, interactive manner, differing from tools like compilers that process static code in a single pass.
Typically, LSP servers are launched by the code editor as background processes to analyze the opened files.
The content of the file is sent to the LSP server with a \code{textDocument/didOpen} message.
When a developer interacts with the code editor, the editor translates these actions into LSP requests and sends them to the LSP server to request analysis or information.
For instance, as illustrated in \cref{fig:clangd-crash}, when a developer places the mouse cursor on a token, the editor issues a \code{textDocument/hover} request to the LSP server, including the URI of the source file and the cursor position as parameters.
The LSP server then analyzes the code at the specified location and returns information \eg{type, signature, documentation} about the symbol, which is displayed to the user in a tooltip.

\section{Problem Formulation and Challenges}\label{sec:formulation}

Testing LSP servers presents unique challenges that traditional software testing approaches cannot adequately address.
Unlike most code analysis tools, which typically focus on generating static source code as input, LSP server testing must account for the dynamic and interactive nature of editor-server communication.
An input to an LSP server consists of two tightly coupled components:
\begin{itemize}
	\item \textbf{Source Code:}
	      A source code file as displayed and edited in the code editor, written in the language supported by the target LSP server.
	      It is also known as a text document in the protocol~\cite{link:lsp}.
	      The full content of the source code is sent to the LSP server when the editor opens the file.
	\item \textbf{Editor Operations:}
	      A sequence of user-driven actions performed on the text document, such as auto-completion, code navigation, mouse hovering, and formatting.
	      These operations are encoded as LSP requests or notifications~\cite{link:lsp} and transmitted to the LSP server in real time.
\end{itemize}
The input space is not limited to source code; it also includes sequences of editor operations that are tightly coupled with the source code.
This coupling between input components introduces unique challenges in LSP testing that require specialized approaches for effective bug discovery.

\subsection{Challenge 1: Combinatorial Input Constraints}
To reveal bugs in LSP servers, test cases must first pass complex input validation layers to reach the core functionality code.
Without satisfying these validations, potential bugs in the server's internal logic remain unexposed~\cite{DBLP:conf/sp/WangCWL17}.
Valid LSP server inputs must satisfy a set of sophisticated and interdependent constraints as specified in the protocol.
Moreover, \emph{these constraints exhibit complex combinatorial properties}.
They apply not only to each source code or editor operation independently, but also to specific combinations of the two components interacting together.
For example, the parameters of an auto-completion operation, a hover operation (as in \cref{fig:clangd-crash}), or a definition operation must refer to a valid position within the source code.

Without systematic awareness of these constraints, traditional fuzzing approaches such as binary~\cite{DBLP:conf/woot/MaierEFH20} or grammar-based~\cite{DBLP:conf/ndss/AschermannFHJST19,DBLP:conf/issta/SrivastavaP21,DBLP:conf/sigsoft/SteinhofelZ22,DBLP:journals/pacmse/AmayaSZ25} fuzzers are unlikely to generate valid editor operations that can satisfy the combinatorial constraints and thus reach the core LSP server functionality.
For instance, while a grammar-based fuzzer can easily generate an input with all the required fields to bypass the request parsing component, the request may point to a meaningless position and thus be rejected when the LSP server tries to process it.

\subsection{Challenge 2: Two-Dimensional Diversity Requirements}\label{sec:challenge-2}

While addressing protocol constraints enables valid test cases, effective bug discovery further requires exploring diverse combinations of inputs.
Although diversity is a common requirement for software testing techniques~\cite{DBLP:conf/pldi/YangCER11,DBLP:journals/tosem/DutraGZ24,DBLP:conf/issta/SrivastavaP21}, LSP server testing presents a unique two-dimensional diversity requirement for test cases.
Its effectiveness depends on the interplay between two distinct dimensions of diversity:
\begin{itemize}
	\item \textbf{Source Code Diversity:}
	      The structural variety within the source code, including both well-formed and incomplete or invalid code fragments.
	\item \textbf{Editor Operation Diversity:}
	      The range of editor operations performed at various code constructs within the source code file.
\end{itemize}

A key aspect of source code diversity is the frequent presence of incomplete or invalid code during interactive editing.
For example, when a developer writes an incomplete variable declaration, the LSP server is expected to offer it as a completion candidate, demonstrating its ability to perform \emph{partial analysis} on ill-formed code.
Such scenarios can trigger unique behaviors and error-handling routines, further expanding the exploration space for LSP server testing.

Editor operation diversity is another important dimension, which is linked to the dynamic and interactive nature of LSP servers.
An editor operation is \emph{location-sensitive}: it can target a specific code symbol, a selection region, the entire code file, or the whole project.
This \emph{referential} relationship between source code and editor operations results in a vast space of possibilities: the triggered behavior depends not only on the operation itself but also on the particular code construct(s) it references.
The same operation can exercise entirely different logic paths depending on where it is applied.
For instance, a hover operation at a function call typically prompts the LSP server to extract its documentation, whereas the same request at a variable declaration leads the LSP server to infer its type.
The bug illustrated in \cref{fig:clangd-crash} exemplifies this interplay: it could only be triggered by the specific combination of source code containing a variable initialization with an incomplete function call and a hover request at the equal sign.
Neither component alone would have exposed this bug.

Therefore, the exploration space formed by these two dimensions is not merely additive but multiplicative, which spans all the possible combinations of source code, selection targets, and editor operation sequences.
Effective LSP server testing must systematically explore combinations of source code and editor operations, including those involving incomplete or invalid code, to efficiently exercise diverse code paths and error-handling routines.
This combinatorial challenge necessitates fuzzing approaches specifically designed to navigate the unique input space of LSP servers.

\section{Approach: \lspfuzz{}}\label{sec:approach}
\begin{figure*}[t]
	\centering
	\includegraphics[width=\linewidth]{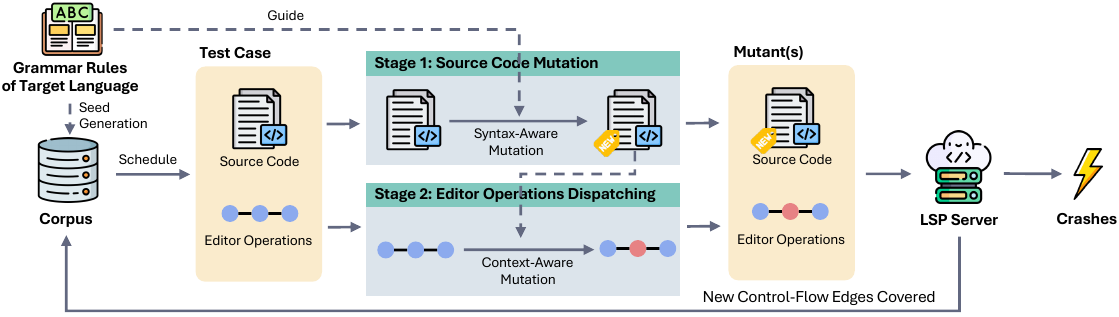}
	\caption{Overview of \lspfuzz{}}
	\label{fig:lsp-fuzz-overview}
\end{figure*}

The two challenges in LSP testing necessitate a specialized testing technique that systematically addresses both the combinatorial input constraints and the two-dimensional diversity requirements.
In this section, we present our approach, \lspfuzz{}, a grey-box mutational fuzzer designed to detect crashes and memory corruptions in LSP server implementations.

\cref{fig:lsp-fuzz-overview} shows an overview of \lspfuzz{}.
The fuzzing process begins by initializing the corpus with a set of randomly generated source code using \textsc{Tree-Sitter}~\cite{software:tree-sitter} grammar rules of the target language.
For each fuzzing iteration, the scheduler selects a seed from the corpus, which then undergoes a two-stage mutation pipeline.
The mutated test case is run on the target LSP server, and its runtime behavior is monitored.
If the test case triggers a previously unexplored control-flow edge, it is added to the corpus for further mutation.
When execution results in a crash with a unique stack trace, the test case is recorded as a potential bug.
This process continues iteratively until the allocated time budget is exhausted.

At the core of \lspfuzz{} is a two-stage mutation pipeline, which enables the generation of test cases with two interconnected parts \ie{source code and editor operations}.
This two-stage pipeline addresses the sophisticated constraints of LSP.
The first stage performs syntax-aware mutation to produce source code with diverse syntactic combinations, triggering the code analysis logic for various types of code.
The second stage generates editor operations targeting the mutated source code produced by the first stage, applying various strategies to dispatch editor operations that are likely to trigger deeper code analysis behaviors in LSP servers.
Together, these strategies exercise various functionalities of LSP servers by producing diverse inputs.
The following subsections detail this pipeline.

\subsection{Stage I: Source Code Mutation}\label{sec:code-mutation}

The first stage in the mutation pipeline aims to produce source code with diverse syntactic features.
Such code serves as a foundation for exploring LSP server behaviors.
We combine formal grammar-based generation with real-world code patterns to achieve both structural and semantic diversity.

\subsubsection{Grammar-Based Mutation}\label{sec:syn-mut}

For source code mutation, we leverage the random tree mutation strategy that has been proven effective in many grammar-based fuzzers~\cite{DBLP:conf/ndss/AschermannFHJST19,DBLP:conf/issta/SrivastavaP21,DBLP:journals/pacmse/AmayaSZ25}.
Specifically, we randomly select a non-terminal node in the derivation tree of the source code and replace it with a new sub-tree rooted at the current non-terminal node type.
To generate the replacement node, we randomly pick a production rule for the current non-terminal and expand it recursively.
This ensures that the generated code adheres to the grammar rules of the target language and allows us to explore a wide range of possible syntactic combinations in that language.

While leveraging grammar rules ensures syntactic correctness, the generated code often lacks the semantic richness and complexity of real-world code~\cite{DBLP:journals/pacmpl/LiTS24}.
To enrich the generated code and trigger various code analysis logic in the LSP server, we incorporate real-world code as ingredients for source code mutation, adopting an idea similar to \textsc{LangFuzz}~\cite{DBLP:conf/uss/HollerHZ12}.
Specifically, when generating the replacement node, we randomly choose a sub-tree from a code fragment pool (\cref{sec:impl}) with a compatible node type.

These two strategies enable us to produce diverse uses of the source code in the target language, which serves as a strong basis for exploring LSP server behaviors.

\subsubsection{Invalid Code Injection}\label{sec:invalid-code-injection}
LSP servers must frequently process invalid code, as developers often produce incomplete or incorrect code during editing.
Supporting partial analysis on such code is essential for LSP servers and exposes unique behaviors relevant for testing.
For example, an LSP server should recognize an incomplete variable declaration and offer it as a completion candidate.
Thus, generating invalid code is necessary to explore the input space and ensure the robustness of LSP servers.

\begin{figure}[t]
	\centering
	\includegraphics[width=0.95\linewidth]{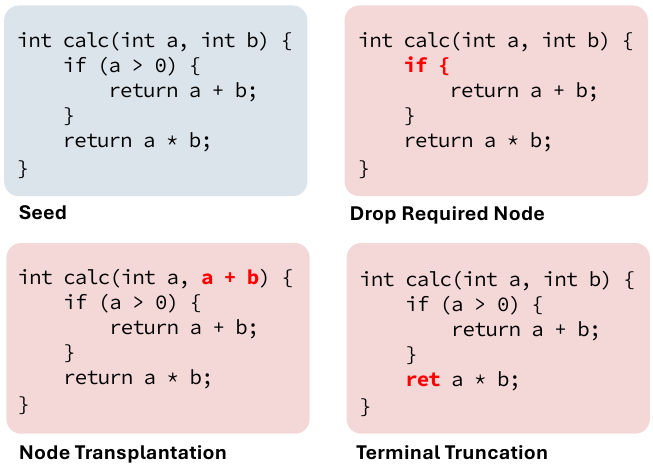}
	\caption{Examples of Mutation Operations Leading to Invalid Code}
	\label{fig:invalid-code-mutations}
\end{figure}

To enable \lspfuzz{} to explore this region in the input space, we produce invalid code at a controlled frequency.
The generation of invalid code must be carefully controlled to achieve two competing objectives:
\begin{enumerate*}
	\item The code should differ from normal valid code sufficiently to trigger the server's partial analysis mechanisms, and
	\item It should not be so malformed that it is immediately rejected without further processing.
\end{enumerate*}
To balance these objectives, we designed three mutation operators (illustrated in \cref{fig:invalid-code-mutations}) that produce code with a generally well-formed structure but with localized errors or misalignments.
In addition, they emulate common scenarios where developers produce intermediate code during programming.
\begin{itemize}
	\item \textbf{Drop Required Nodes}: Removes required child nodes from a non-terminal in the parse tree.
	      For example, omitting the condition of an \code{if} statement creates syntactically incomplete yet structurally meaningful code.
	      It simulates scenarios where developers are creating a code construct \eg{classes, functions, etc.}.
	      The resulting code can trigger the partial analysis mechanism, whereby LSP servers recognize and process incomplete code constructs.

	\item \textbf{Node Transplantation}: Replaces a node with another of a different type, such as substituting a parameter declaration with an expression node.
	      This simulates scenarios where developers are refactoring or reorganizing code.
	      This operator helps evaluate whether LSP servers can robustly process code with localized structural inconsistencies that may arise during such changes.

	\item \textbf{Terminal Truncation}: Truncates terminal tokens, such as cutting \code{return} to \code{ret} or leaving a string literal without a closing quote.
	      This simulates scenarios where developers are typing or deleting a code symbol, testing LSP servers' ability to process partial symbols.

\end{itemize}

The generation of invalid code complements valid code produced by random tree mutation and enables \lspfuzz{} to explore both the well-formed and ill-formed regions of the input space, mirroring the spectrum of source code states encountered during real-world software development.

Our approach to source code mutation produces diverse test inputs by generating structurally varied code through grammar-based techniques, incorporating real-world patterns and semantics, including both valid and invalid source code.
The resulting rich testing space, when combined with our editor operation dispatching (\cref{sec:editor-ops}), enables us to effectively explore the input space of LSP servers.

\subsection{Stage II: Editor Operations Dispatching}\label{sec:editor-ops}

Given the source code produced in the first stage as context, the second stage of the mutation pipeline dispatches editor operations targeting various constructs in the source code.
Specifically, we randomly select an operation from all the available LSP operations, generate its parameters, and insert it into the editor operation sequence.
Among all the parameters, the code construct targeted by the operation is the predominant factor for triggering behaviors in the LSP server.
Therefore, the key to editor operation dispatching is to find the target construct for the editor operation.
We consider two strategies that focus on the syntactic features of the source code and the responses produced by LSP servers.

\subsubsection{Syntactic-Category-Level Randomization}
In LSP, the code context of an editor operation is specified by line and column, referring to a specific position or range in the source code.
However, trivial randomization may not effectively yield editor operations that trigger diverse LSP server behaviors.
For example, character-level randomization may frequently select positions within the same token (especially for longer ones) or with the same node types (especially for common node types), which typically trigger identical or similar LSP server behaviors.
Instead, we should perform random selection in a space where each element potentially links to distinct LSP server behaviors.

To achieve this goal, we analyze the behaviors of LSP servers given various code and operations to understand how they process these editor operation requests.
We observed that the code construct targeted by an editor operation can affect the behavior of LSP servers when processing it.
Specifically, different syntactic categories of code constructs \eg{function, variable} can lead to significantly different behaviors.
For example, a \code{textDocument/hover} at a function can lead to documentation extraction, while the same operation at a variable can lead to type inference.
Thus, we should project the locations in the source code to a space that reflects the syntactic category of their corresponding code constructs.

Inspired by this observation, we introduce the \emph{syntactic signature} for each node in the parse tree.
Specifically, a syntactic signature of level \(n\) refers to a sequence \(\langle t_0, t_1, \ldots, t_n \rangle\) where \(t_0\) is the type of the node, and \(t_n\) is the type of its \(n\)\textsuperscript{th} parent node.
A syntactic signature captures not only the syntactic characteristic of the node itself, but also its contextual information.
This is crucial because nodes with the same type can have different semantics depending on where they appear in the syntax tree.
For example, in C, a function name and a variable name both have the type \code{identifier}, but their level-1 syntactic signatures differ: \(\langle \text{\code{identifier}}, \text{\code{function\_declarator}}\rangle\) vs.
\(\langle \text{\code{identifier}}, \text{\code{init\_declarator}}\rangle\).
They can lead to different LSP server behaviors when targeted by editor operations \eg{hover} as mentioned earlier.
Syntactic signature works at the grammar level and does not rely on language-specific knowledge, which allows us to categorize code constructs in an arbitrary language.

When identifying target code constructs to dispatch editor operations, we perform random selection at the syntactic-category level.
Specifically, we group the locations in the source code based on the syntactic signature of their corresponding nodes.
Each group contains the locations with parse tree nodes in a specific syntactic category.
Then we dispatch editor operations by randomly choosing code constructs from each group.
Such syntactic-category-level randomization enables us to distribute exploration among the potential trigger behaviors of LSP servers and reduces the chance of generating redundant test cases.

\subsubsection{Editor Operations Enrichment}

Certain symbols in the source code lead to deeper code analysis, such as the semantic-level symbols recognized by LSP servers and code with diagnostic information.
However, this kind of information cannot be derived from syntax-level analysis alone.
To better trigger diverse behaviors linked to such symbols, we design two strategies to identify these symbols and prioritize dispatching editor operations that target them.

\paragraph{Semantic Symbols}
Semantic symbols are a set of predefined code constructs \eg{functions, types, and fields} recognized by the target LSP servers.
These are regarded as well-formed code constructs because they pass basic validation and are present in the symbol table of the LSP server.
Dispatching editor operations on them is more likely to trigger deeper code analysis within the LSP server.
We identify them by capturing the responses generated by the LSP server for \code{textDocument/documentSymbol} and \code{workspace/symbol}, which contain the locations of these semantic symbols in the source code.
During mutation, we prioritize dispatching editor operations to these locations.

\paragraph{Diagnostic Information}
Code constructs containing localized errors, whether syntactic or semantic, can trigger partial analysis logic in LSP servers (\cref{sec:challenge-2}).
Apart from syntax errors we inject during source code mutation, semantic errors are an important counterpart that can lead to partial analysis, and we leverage the diagnostic information generated by LSP servers to identify them.
Specifically, the LSP server can emit detected errors in the source code by sending \code{textDocument/publishDiagnostics} requests to the code editor.
We capture these requests and prioritize dispatching editor operations targeting the nodes that contain a child with a diagnostic.
These nodes contain localized errors and are likely to trigger more partial analysis logic in the LSP servers.

\paragraph{Inter-Operation Dependencies}
In LSP, several editor operations depend on the outcome of other operations.
For example, a \code{callHierarchy/outgoingCalls} operation requires a \code{CallHierarchyItem} in its parameter, which is generated by the \code{textDocument/prepareCallHierarchy} operation.
For such editor operations, we save LSP server responses for each seed in the corpus and use the associated response data to construct the parameters of these operations during mutation.

These two strategies work together to enrich our editor operations, directing fuzzing efforts toward code constructs that are more likely to exercise the deeper analysis logic of LSP servers and uncover potential bugs in their implementation.

\subsection{Implementation}\label{sec:impl}

\paragraph{Overall Framework}
To facilitate future extension and enable seamless interoperability with different fuzzing strategies, we implemented \lspfuzz{} based on \textsc{LibAFL}~\cite{DBLP:conf/ccs/FioraldiMZB22} with \locLspfuzz{} lines of Rust code, leveraging its modular architecture.
We implemented custom mutators for our two-stage mutation pipeline while utilizing \textsc{LibAFL}'s built-in coverage tracking and scheduling modules.

\paragraph{Source Code Processing}
For source code parsing, mutation, and code context identification, we integrated \textsc{Tree-Sitter}~\cite{software:tree-sitter}, a parsing system that supports over 100 programming languages.
This integration provides \lspfuzz{} with the capability to test LSP servers across diverse programming languages without requiring language-specific tooling.

\paragraph{Code Fragment Pool}
Before the fuzzing campaign, we construct the code fragment pool (\cref{sec:syn-mut}) by parsing example files and test cases in the code repository of the target LSP server, which contains code in the target language of the LSP server.
Users of \lspfuzz{} may optionally enrich it with additional source files \eg{mined from GitHub}.
The probability of selecting a real-world code fragment during mutation is controlled by a configurable hyperparameter, which is set to 0.2 by default.

\paragraph{Fuzz Targets}
\lspfuzz{} is compatible with the fuzz targets for \textsc{AFL++}~\cite{DBLP:conf/woot/MaierEFH20} and \textsc{libFuzzer}~\cite{software:libfuzzer}.
To prepare a fuzz target, users should modify the entry point \ie{the \code{main} function} of the target LSP server to read inputs from an in-memory byte array instead of \code{stdin}~\ie{shared-memory fuzzing}.
The fuzz target should be instrumented for collecting coverage feedback.
Optionally, users can remove code unrelated to LSP functionality \eg{CLI option parsing, configuration file loading} to achieve higher fuzzing throughput~\cite{DBLP:conf/ccs/KleesRCW018}.

\section{Evaluation}\label{sec:evaluation}

We evaluate \lspfuzz{} through four research questions to assess its effectiveness and usefulness.
Specifically, we aim to answer the following research questions.

\begin{itemize}
	\item \textbf{RQ1:}
	      Is \lspfuzz{} effective in detecting crashes in LSP servers?
	\item \textbf{RQ2:}
	      Can \lspfuzz{} outperform baselines in code coverage and crash detection?
	\item \textbf{RQ3:}
	      Is the two-stage mutation pipeline essential for the effectiveness of \lspfuzz{}?
	\item \textbf{RQ4:}
	      Is \lspfuzz{} useful in finding previously unknown bugs in real-world LSP servers?
\end{itemize}

These research questions assess the effectiveness and practical utility of \lspfuzz{}.
RQ1 focuses on whether \lspfuzz{} is effective at detecting crashes in LSP servers, using standard fuzzing metrics such as the number of detected crashes and code coverage.
RQ2 compares \lspfuzz{} against baseline fuzzing approaches to determine if it can achieve higher code coverage and detect more crashes.
RQ3 investigates the necessity and impact of the two-stage mutation pipeline by comparing \lspfuzz{} to a variant without this feature.
RQ4 evaluates the practical usefulness of \lspfuzz{} by examining its ability to uncover previously unknown bugs in production LSP servers and the responses from development teams.

\subsection{Subject LSP Servers}\label{sec:subjects}

\begin{table}
	\caption{Subject LSP Servers for Evaluation}\label{tab:subjects}
	\begin{threeparttable}
	\begin{tabularx}{\linewidth}{LlLLR}
		\toprule
		\textbf{Name}     &
		\textbf{Vendor}   &
		\textbf{Language} &
		\textbf{Version}  &
		\textbf{Popularity\textsuperscript{*}}
		\\
		\midrule
		\texttt{clangd}   & LLVM           & C/C++    & v20.1.4    & 1.8M \\
		\texttt{sorbet}   & Stripe Inc.    & Ruby     & v0.5.11031 & 901k \\
		\texttt{verible}  & CHIPS Alliance & Verilog  & v0.0.3157  & 1.1M \\
		\texttt{solc}     & Ethereum       & Solidity & v0.8.29    & 1.6M \\
		\bottomrule
	\end{tabularx}
	\begin{tablenotes}
		\item \scriptsize{} *Number of installs of the associated \textsc{VSCode} extension at the time of writing
	\end{tablenotes}
\end{threeparttable}

	\vspace{-1em}
\end{table}
We evaluated \lspfuzz{} by selecting LSP servers from the top 200 most installed \textsc{VSCode} programming language extensions using three criteria:
\begin{enumerate*}
	\item they are actively maintained \ie{have had a release in the past three months},
	\item they are open-source, allowing us to modify the source code to prepare fuzz targets, and
	\item they are compatible with AFL++~\cite{DBLP:conf/woot/MaierEFH20} instrumentation.
\end{enumerate*}
\cref{tab:subjects} lists the selected LSP servers, which are the top-4 most popular LSP servers from the candidates.
They are designed for diverse languages and ecosystems: \texttt{clangd} is part of LLVM, providing language services for C/C++; the Solidity compiler (\texttt{solc}) includes LSP support for Ethereum's smart contracts; \texttt{sorbet} is Stripe's LSP server for Ruby; and \texttt{verible} supports System Verilog development.
These LSP servers are large-scale, production-grade, mature software.
They share similar complexity with compilers.
Their associated \textsc{VSCode} extensions have attracted thousands to millions of installs, accounting for 96\% of the total installations among the candidates.

We prepared fuzz targets following the approach presented in \cref{sec:impl}.
To achieve higher throughput, we removed the code for CLI option parsing and configuration file loading, as they are irrelevant to LSP.
We enabled \code{AddressSanitizer}~\cite{link:asan} during compilation to detect memory corruptions.
All four LSP servers are written in C/C++, and we instrumented the fuzz targets with the LLVM LTO mode provided by AFL++~\cite{DBLP:conf/woot/MaierEFH20}.

\subsection{Baselines and Experiment Setup}\label{sec:baseline-setup}

To the best of our knowledge, no existing technique specifically targets LSP server testing.
Therefore, we implemented three baseline approaches using \textsc{LibAFL}~\cite{DBLP:conf/ccs/FioraldiMZB22}.

\paragraph{\binaryBaseline{}}
To understand the advantage of \lspfuzz{} over current practice, we created a baseline, \binaryBaseline{}, using the binary mutators from \textsc{LibAFL}~\cite{DBLP:conf/ccs/FioraldiMZB22}, which share the same algorithm as AFL++~\cite{DBLP:conf/woot/MaierEFH20}.
\binaryBaseline{} is inspired by the practice that \texttt{clangd} and \texttt{sorbet} use binary fuzzers in their testing workflows, as evidenced in their public repositories.
As a result, \binaryBaseline{} represents the current state of practice in fuzzing LSP servers and provides a comparison point to demonstrate the advantage of \lspfuzz{} over current practice.

\paragraph{\grammarBaseline{}}
We argued in~\cref{sec:formulation} that a grammar-based fuzzer with LSP schema would be ineffective without considering the complex semantic relationships between source code and LSP operations.
To validate this argument, we designed a baseline, \grammarBaseline{}, that implements a pure grammar-based approach.
It leverages the grammar mutators built into \textsc{LibAFL}~\cite{DBLP:conf/ccs/FioraldiMZB22} and operates with a grammar constructed from the LSP schema, which specifies the syntactic structure of each LSP message type.
Comparing \lspfuzz{} against \grammarBaseline{} helps empirically demonstrate the advantage of maintaining semantic consistency in LSP server testing.
In addition, ten randomly sampled inputs generated by \grammarBaseline{} are used as initial seeds for \binaryBaseline{}.

\paragraph{\twoDimBaseline{}}
As discussed in \cref{sec:formulation}, LSP server inputs consist of two dimensions: source code and editor operations.
To capture this structure, we adapted a two-dimensional baseline from the \code{MultipartInput} in \textsc{LibAFL}.
The adaptation was done following the LSP specifications~\cite{link:lsp}.
Specifically, this baseline generates code and editor operations using the built-in mutation operators in \textsc{LibAFL}.
The generated code is first sent to the LSP server with a \code{textDocument/didOpen} message, followed by a sequence of editor operations.
\twoDimBaseline{} generates inputs that are guaranteed to contain both code and editor operations.
Comparing \lspfuzz{} with \twoDimBaseline{} helps to illustrate the benefits of our systematic mutation strategy tailored for LSP server testing.

Compared with \lspfuzz{} these baselines either ignore the structural and semantic relationships in LSP inputs or fail to model the interaction between source code and editor operations, resulting in test cases that lack meaningful correspondence to real LSP usage.

Although we considered LLM-based approaches, they do not support LSP servers and adapting them is non-trivial.
For example, \textsc{ChatAFL}~\cite{DBLP:conf/ndss/MengMBR24} fails to start when LSP is specified as the protocol.
We also considered existing multi-dimensional fuzzing approaches.
However, they either
\begin{enumerate*}
	\item do not intent to address the challenges in LSP server testing, or
	\item cannot be adapted to LSP due to their reliance on target-specific properties.
\end{enumerate*}
Thus, we did not include them as baselines.
We discuss these approaches in \cref{sec:related-work}.

All experiments were conducted on a dual AMD EPYC 7773X machine (128 cores, 1TB RAM, AlmaLinux 9.5).
Each fuzzing process used one core for 24 hours, repeated ten times, totaling 200 CPU-days.

\subsection{RQ1: Effectiveness of \lspfuzz{} in Crash Detection}
\begin{table*}[t]
	\caption{Reproducible Crashes with Unique Stack Traces (RQ1 -- RQ3)}\label{tab:crashes-st}
	\begin{threeparttable}
	\newcommand*{\tot}[0]{\textbf{Total}}
	\newcommand*{\avg}[0]{\textbf{Avg}}
	\newcommand*{\cl}[0]{\textbf{Code}}
	\newcommand*{\ot}[0]{\textbf{EOP}}
	\setlength{\tabcolsep}{3pt}
	\begin{tabularx}{\linewidth}{l|RRRR|RRRR|RRRR|RRRR|RRRR}
		\toprule
		\textbf{Subject}                      &
		\multicolumn{4}{c|}{\texttt{clangd}}  &
		\multicolumn{4}{c|}{\texttt{sorbet}}  &
		\multicolumn{4}{c|}{\texttt{verible}} &
		\multicolumn{4}{c|}{\texttt{solc}}    &
		\multicolumn{4}{c}{\textbf{All}}
		\\
		\midrule
		\textbf{Approach}\textsuperscript{1}  & \tot & \cl & \ot & \avg & \tot & \cl & \ot & \avg & \tot & \cl & \ot & \avg & \tot & \cl & \ot & \avg & \tot & \cl & \ot & \avg
		\\
		\midrule
		\lspfuzz{}                            & 499  & 483 & 16  & 70.9 & 64   & 27  & 37  & 28.5 & 399  & 20  & 379 & 74.6 & 29   & 25  & 4   & 7.1  & 991  & 555 & 436 & 181.1 \\
		\binaryBaseline{}                     & 0    & -   & -   & 0.0  & 2    & -   & -   & 2.0  & 2    & -   & -   & 2.0  & 0    & -   & -   & 0.0  & 4    & -   & -   & 4.0   \\
		\grammarBaseline{}                    & 5    & 5   & 0   & 2.3  & 2    & 2   & 0   & 2.0  & 5    & 5   & 0   & 2.2  & 0    & 0   & 0   & 0.0  & 12   & 12  & 0   & 6.5   \\
		\twoDimBaseline{}                     & 122  & 122 & 0   & 14.9 & 7    & 7   & 0   & 2.8  & 7    & 7   & 0   & 3.1  & 1    & 1   & 0   & 1.0  & 137  & 137 & 0   & 21.8  \\
		\lspfuzznc{}                          & 437  & 437 & 0   & 64.4 & 25   & 25  & 0   & 14.6 & 251  & 21  & 230 & 41.5 & 21   & 21  & 0   & 2.7  & 734  & 504 & 230 & 120.2 \\
		\bottomrule
	\end{tabularx}
	\begin{tablenotes}
		\scriptsize
		\item 1.
		Total are all the crashes, Code are the code-loading crashes, EOP are operation-triggered crashes, Avg are the average number of crashes over the 10 repetitions.
		\item 2.
		Test cases from \binaryBaseline{} may not be valid LSP inputs and cannot always be classified as code-loading or operation-triggered crashes.
	\end{tablenotes}
\end{threeparttable}

	\vspace{-1.2em}
\end{table*}

RQ1 evaluates the effectiveness of \lspfuzz{} by the number of detected crashes, which is a typical evaluation metric for fuzzers~\cite{DBLP:conf/ccs/KleesRCW018}.
We deduplicated the crashes with stack hashes following the practice of existing work~\cite{DBLP:conf/ccs/KleesRCW018}, and report the results in \cref{tab:crashes-st}.
In our subject LSP servers, \lspfuzz{} detected a total of 991 crashes in 10 runs of 24 hours, with an average of 181.1 per run.

To better understand how these crashes are triggered, we categorized them into two types based on the stage at which they occur, reflecting the complexity required to trigger them.
\begin{itemize}
	\item \textbf{Code-Loading Crashes:}
	      These occur when the LSP server receives and analyzes source code from the editor.
	      Loading source code alone can trigger these crashes, typically during code parsing or initial analysis, without requiring specific editor interactions.
	\item \textbf{Operation-Triggered Crashes:}
	      These occur when the LSP server processes specific editor operations on certain source code.
	      They require particular combinations of source code and editor operations, and typically manifest in components handling language features or editor functionalities.
\end{itemize}

The columns \emph{Code} and \emph{EOP} in \cref{tab:crashes-st} show the breakdown of the crashes into these two categories, respectively.
The number of code-loading crashes is large, especially for \texttt{clangd}, which has a huge amount of code for various C/C++ analyses.
This is as expected because code-loading crashes are easier to trigger compared with operation-triggered crashes, as they do not require specific combinations of source code and editor operations.
Nevertheless, \lspfuzz{} is effective at discovering operation-triggered crashes, with 44\% of the detected crashes being operation-triggered.
Such crashes require specific combinations of source code and editor operations.

The results highlight the significant crash detection capabilities of \lspfuzz{}.
In the following, we show two cases to illustrate the crashes found by \lspfuzz{}.
These two cases highlight different strengths of our approach.

\paragraph{Case Study I: Crash in \texttt{clangd} on Formatting}
This case demonstrates the effectiveness of our invalid code injection mechanism (\cref{sec:invalid-code-injection}).
\lspfuzz{} found a crash in the C/C++ LSP server \texttt{clangd}, which is triggered by the combination of a code snippet with localized errors and a \code{textDocument/formatting} operation.
The code snippet contains a C++ \code{decltype} specifier missing its closing parenthesis.
This simulates a realistic scenario where a developer is writing a \code{decltype} specifier but has not yet finished.
When processing the formatting operation, the formatter in \texttt{clangd} attempted to locate the end of the \code{decltype} sequence but encountered a \code{null} pointer, resulting in a segmentation fault.
The malformed \code{decltype} specifier was produced by our mutation operators for injecting invalid code, which revealed this bug when combined with a formatting editor operation.
Without invalid code injection, \lspfuzz{} could not produce such a test case.
The bug affects users in common programming workflows, as code formatting is a frequently used feature during development.
LLVM developers confirmed and fixed this bug, which had remained hidden in \texttt{clangd} for two years.

\paragraph{Case Study II: Crash in \texttt{sorbet} on Go to Implementation}
This case demonstrates the effectiveness of our editor operation dispatching strategies (\cref{sec:editor-ops}).
\lspfuzz{} discovered a crash in the Ruby LSP server \texttt{sorbet} triggered by the combination of a malformed lambda expression \code{->...\{\}} and a \code{textDocument/implementation} operation specifically targeting the arrow symbol.
This combination resulted in an assertion violation in \texttt{sorbet}, causing the LSP server to abort.
This case exemplifies the power of our syntactic signature approach: although the arrow of a lambda expression spans only two characters in the source code, our system recognized its syntactic significance as an \code{arrow} within a \code{lambda} in the Ruby grammar rules, and balanced the chance to interact with it and other long units like a lengthy literal.
Furthermore, the triple dot operator between the arrow and its body triggered diagnostics, causing it to be prioritized by our editor operation enrichment mechanism.
This case highlights how our approach can identify interesting positions within the code for applying editor operations, even for small syntax elements.
The \texttt{sorbet} development team confirmed and fixed this bug, which had remained undetected in the project for four years.

\begin{table}[t]
	\caption{Reproducible Crashes at Unique Locations (RQ1 -- RQ4)}\label{tab:crashes-cl}
	\begin{threeparttable}
	\newcommand*{\tot}[0]{\textbf{Total}}
	\newcommand*{\avg}[0]{\textbf{Avg}}
	\setlength{\tabcolsep}{3pt}
	\begin{tabularx}{\linewidth}{l|RR|RR|RR|RR|RR}
		\toprule
		\textbf{Subject}                      &
		\multicolumn{2}{c|}{\texttt{clangd}}  &
		\multicolumn{2}{c|}{\texttt{sorbet}}  &
		\multicolumn{2}{c|}{\texttt{verible}} &
		\multicolumn{2}{c|}{\texttt{solc}}    &
		\multicolumn{2}{c}{\textbf{All}}
		\\
		\midrule
		\textbf{Approach}\textsubscript{1}    & \tot & \avg & \tot & \avg & \tot & \avg & \tot & \avg & \tot & \avg
		\\
		\midrule
		\lspfuzz{}                            & 60   & 14.3 & 32   & 19.8 & 19   & 16.1 & 4    & 4    & 115  & 54.2 \\
		\binaryBaseline{}                     & 0    & 0.0  & 2    & 2.0  & 2    & 2.0  & 0    & 0    & 4    & 4.0  \\
		\grammarBaseline{}                    & 5    & 2.3  & 2    & 2.0  & 4    & 1.8  & 0    & 0    & 11   & 6.1  \\
		\twoDimBaseline{}                     & 5    & 2.7  & 7    & 2.8  & 3    & 2.4  & 1    & 1.0  & 16   & 8.9  \\
		\lspfuzznc{}                          & 34   & 9.9  & 15   & 9.7  & 13   & 10.1 & 1    & 1    & 62   & 30.6 \\
		\bottomrule
	\end{tabularx}

	\begin{tablenotes}
		\scriptsize
		\item 1.
		Total are all the crashes, Avg are the average number of crashes over the 10 repetitions.
	\end{tablenotes}
\end{threeparttable}

	\vspace{-0.75em}
\end{table}

Last but not least, while stack-trace-based crash deduplication is a common strategy~\cite{DBLP:conf/ccs/KleesRCW018}, for LSP servers, crashes with different stack traces may not always correspond to distinct bugs.
This is because the same underlying fault can be triggered in multiple contexts \eg{a bug in the type inference module can be triggered by both hover and go-to-type-definition}, resulting in different stack traces depending on the execution path.
Therefore, we further deduplicated the crashes in \cref{tab:crashes-st} by the program locations where they occur, and report the results in \cref{tab:crashes-cl}.

Even with deduplication at this level, \lspfuzz{} can still detect an average of 54.2 crashes across 10 runs, with 14.3 in \texttt{clangd}, 19.8 in \texttt{sorbet}, 16.1 in \texttt{verible}, and 4 in \texttt{solc}.

\begin{summary}
	\lspfuzz{} is effective in detecting crashes in real-world LSP servers.
	On average, it detected 181.1 crashes with distinct stack traces and 54.2 crashes at different program locations across four LSP servers over 10 runs.
	Over 40\% of the detected crashes manifest with specific combinations of source and editor operations, which are difficult to trigger.
\end{summary}

\subsection{RQ2: Comparison with Baselines}\label{sec:rq2}

RQ2 aims to investigate whether \lspfuzz{} can outperform baselines in two typical metrics for evaluating fuzzers: edge coverage \ie{number of control-flow edges covered} and number of crashes detected~\cite{DBLP:conf/ccs/KleesRCW018}.

\begin{table}[t]
	\caption{Average Edge Coverage Achieved in 24 Hours (RQ2 and RQ3)}\label{tab:coverage}
	\begin{tabularx}{\linewidth}{LLrr}
	\toprule
	\textbf{Subject}                  & \textbf{Approach}  & \textbf{Edge Coverage \textpm{} 95\% CI} & \textbf{\% \lspfuzz{}} \\
	\midrule
	\multirow{5}{*}{\texttt{clangd}}  & \lspfuzz{}         & 348,741 \textpm{} 11,402                 & 100\%                  \\
	                                  & \binaryBaseline{}  & 2,503 \textpm{} 9                        & 0.7\%                  \\
	                                  & \grammarBaseline{} & 71,195 \textpm{} 1,915                   & 20.4\%                 \\
	                                  & \twoDimBaseline{}  & 90,450 \textpm{} 1,484                   & 25.9\%                 \\
	                                  & \lspfuzznc{}       & 280,966 \textpm{} 8,581                  & 80.6\%                 \\
	\midrule

	\multirow{5}{*}{\texttt{sorbet}}  & \lspfuzz{}         & 83,151 \textpm{} 127                     & 100\%                  \\
	                                  & \binaryBaseline{}  & 6,932 \textpm{} 63                       & 8.3\%                  \\
	                                  & \grammarBaseline{} & 8,302 \textpm{} 2                        & 9.9\%                  \\
	                                  & \twoDimBaseline{}  & 32,755 \textpm{} 149                     & 39.2\%                 \\
	                                  & \lspfuzznc{}       & 73,812 \textpm{} 153                     & 88.4\%                 \\
	\midrule
	\multirow{5}{*}{\texttt{verible}} & \lspfuzz{}         & 29,133 \textpm{} 56                      & 100\%                  \\
	                                  & \binaryBaseline{}  & 8,990 \textpm{} 332                      & 30.9\%                 \\
	                                  & \grammarBaseline{} & 11,874 \textpm{} 290                     & 40.8\%                 \\
	                                  & \twoDimBaseline{}  & 13,504 \textpm{} 318                     & 46.4\%                 \\
	                                  & \lspfuzznc{}       & 20,336 \textpm{} 31                      & 69.8\%                 \\
	\midrule
	\multirow{5}{*}{\texttt{solc}}    & \lspfuzz{}         & 53,720 \textpm{} 72                      & 100\%                  \\
	                                  & \binaryBaseline{}  & 2,816 \textpm{} 14                       & 5.2\%                  \\
	                                  & \grammarBaseline{} & 3,572 \textpm{} 7                        & 6.6\%                  \\
	                                  & \twoDimBaseline{}  & 10,838 \textpm{} 154                     & 20.2\%                 \\
	                                  & \lspfuzznc{}       & 52,987 \textpm{} 97                      & 98.6\%                 \\
	\bottomrule
\end{tabularx}

	\vspace{-1.2em}
\end{table}

\begin{figure*}[t]
	\includegraphics[width=\textwidth]{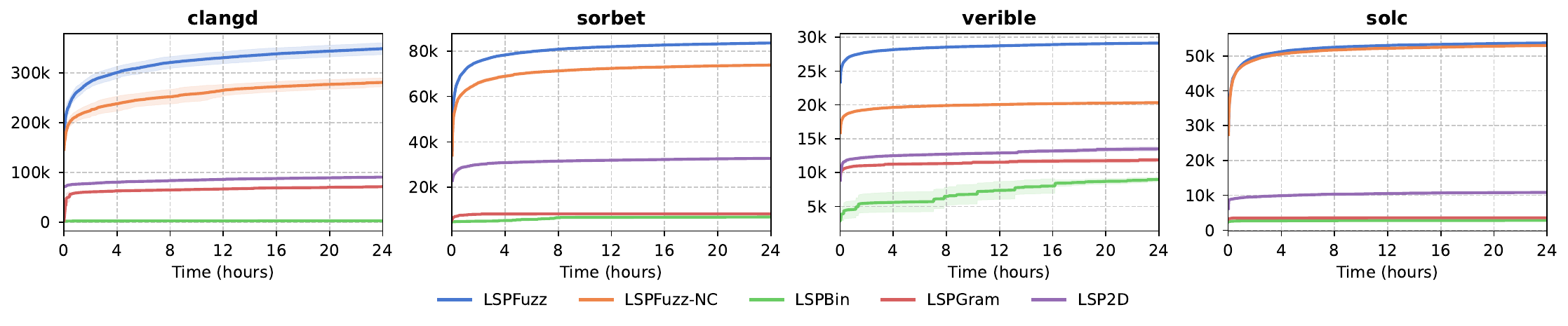}
	\caption{Average Edge Coverage Over Time with 95\% Confidence Interval (RQ2 and RQ3)} \label{fig:coverage} \vspace{-0.75em}
\end{figure*}

\cref{tab:coverage} shows the average number of control-flow edges covered by \lspfuzz{} and the baselines, with a 95\% confidence interval.
Their changes over time are illustrated in \cref{fig:coverage}.
\lspfuzz{} achieved significantly higher edge coverage than the baselines.
For \binaryBaseline{}, which is the status quo in \texttt{clangd} and \texttt{sorbet}, the improvement ranges from 3.2x to 142.9x.
For \grammarBaseline{} and \twoDimBaseline{}, even with the guidance of LSP specification and two dimensions, the improvement is still significant, ranging from 2.2x to 15.1x.

The improvement in code coverage also enables \lspfuzz{} to detect many more crashes compared with the baselines, as shown in \cref{tab:crashes-st}.
On average, \binaryBaseline{} detected 4.0 crashes \grammarBaseline{} detected 6.5 crashes, and \twoDimBaseline{} detected 21.8, which are 2.2\%, 3.6\%, and 12.0\% of \lspfuzz{}, respectively.
\grammarBaseline{} was only able to find code-loading crashes in all subjects and failed to detect any operation-triggered crashes.
This confirms our hypothesis that pure grammar-based approaches struggle with the semantic constraints in LSP.
In addition, even with two input dimensions, \twoDimBaseline{} can find only code-loading crashes.
This demonstrates the advantage of our mutation strategy tailored for LSP servers.
Also, as shown in \cref{tab:crashes-cl}, \lspfuzz{} detected more crashes at unique program locations than the baselines.
The average improvement is 13.6x, 8.9x, and 6.1x for \binaryBaseline{}, \grammarBaseline{}, and \twoDimBaseline{}, respectively.
All four crashes detected by \binaryBaseline{} were in the JSON libraries used for message parsing, not in LSP server logic, and are out of scope of this paper.

The significant improvement can be attributed to our LSP-server-specific problem formulation and mutation strategies.
In comparison, \grammarBaseline{} lacks a two-dimensional formulation, and its generated test cases may not send the source code to LSP servers.
Although \twoDimBaseline{} is built with two-dimensional inputs, it struggles to handle the referential relationship between the code and editor operations, which is brought by the location sensitivity of the editor operations.
As a result, it often produces syntactically correct but semantically meaningless inputs.
Take the input in \cref{fig:clangd-crash} as an example, \twoDimBaseline{} can produce an editor operation that points to nowhere, which violates the location-sensitive constraints of LSP server inputs.
With the two-stage semantic-aware mutation pipeline, \lspfuzz{} can produce more meaningful inputs and more effectively explore the input space.

\begin{summary}
	\lspfuzz{} outperforms our baselines by detecting more (6.1x to 13.6x) crashes, especially for those that are difficult to trigger.
	It also outperforms baselines by a significant margin (2.2x to 142.9x) in edge coverage.
\end{summary}

\subsection{RQ3: Effectiveness of the Two-Stage Mutation Pipeline}

RQ3 investigates the effectiveness of the holistic mutation strategy in \lspfuzz{}, which is the key insight of our approach.
To this end, we construct \lspfuzznc{}, a variant of \lspfuzz{} that mutates the source code and editor operations independently, and run it with the same setup as \lspfuzz{}.

As shown in~\cref{tab:coverage} and~\cref{fig:coverage}, \lspfuzznc{} achieves lower edge coverage than \lspfuzz{}.
In three of our four subjects, \lspfuzznc{} fails to cover many edges reached by \lspfuzz{}.
This is because the lack of context awareness makes our mutation strategies for editor operations ineffective, as these strategies rely on the source code as context.
As a result, \lspfuzznc{} must blindly explore the large, two-dimensional input space.
Such blind exploration often leads to test cases violating LSP constraints \eg{a hover operation pointing at a non-existent location}.
Therefore, \lspfuzznc{} fails to cover as much code as \lspfuzz{} within the same time budget.

The absence of holistic mutation also leads to a decrease in the number of detected crashes, as shown in~\cref{tab:crashes-st}.
These missed crashes are precisely those that are hard to trigger.
As shown in~\cref{tab:crashes-st}, without context awareness, \lspfuzznc{} fails to detect any operation-triggered crashes in three out of four subjects.
This highlights the importance of context awareness in detecting operation-triggered crashes, which are more difficult to trigger compared with code-loading crashes.

An exception is \texttt{solc}; as shown in~\cref{fig:coverage}, the margin between \lspfuzz{} and \lspfuzznc{} is small.
The reason may be that \texttt{solc} provides a very thin layer of LSP features -- it supports only four types of editor operations.
Although most of the covered code is in the underlying code analyzer, \lspfuzz{} can still detect four operation-triggered crashes, demonstrating the effectiveness of our approach even for LSP servers with limited feature sets.
In comparison, for the other three LSP servers that provide rich features, our two-stage mutation pipeline can generate test cases to effectively exercise these features, and thus make a significant difference in coverage.

\begin{summary}
	The two-stage mutation pipeline in \lspfuzz{} is effective.
	It helps achieve better code coverage and detect more crashes, especially the crashes requiring certain combinations of source code and editor operations.
\end{summary}

\subsection{RQ4: Usefulness in Finding Bugs}\label{sec:rq4}

RQ4 aims to reveal the usefulness of \lspfuzz{} in finding bugs in real-world LSP servers.
To evaluate this, we report crashes at unique program locations found by \lspfuzz{} (\cref{tab:crashes-cl}) to the LSP server developers for their confirmation.

We followed ethical guidelines when submitting bug reports.
For bugs with potential security implications, we contacted development teams privately according to their security policies.
To avoid burdening developers~\cite{DBLP:conf/icse/SunLZGS18}, we attempted to minimize test cases by removing irrelevant code and editor operations while preserving the bug-triggering behavior.
Due to the unique LSP test case format, minimization was performed manually, which was labor-intensive and time-consuming.
Therefore, we reported only the crashes detected in the first repetition of our experiments.
However, we have validated that all 115 crashes at unique locations are reproducible.
They are also available in our artifacts~\cite{artifact:zenodo} for readers to examine.

\begin{table}[t]
	\caption{Status of the Reported Bugs (RQ4)}\label{tab:bugs}
	\begin{tabularx}{\linewidth}{l|RRR}
	\toprule
	\textbf{LSP Server} & \textbf{Reported} & \textbf{Confirmed} & \textbf{Fixed} \\
	\midrule
	\texttt{clangd}     & 13                & 13                 & 6              \\
	\texttt{sorbet}     & 20                & 14                 & 13             \\
	\texttt{verible}    & 14                & 12                 & 4              \\
	\texttt{solc}       & 4                 & 3                  & 3              \\
	\midrule
	\textbf{Total}      & \reportedBugs{}   & \confirmedBugs{}   & \fixedBugs{}   \\
	\bottomrule
\end{tabularx}

	\vspace{-1.2em}
\end{table}

\cref{tab:bugs} shows the status of bugs we reported to LSP developers.
In total, we reported \reportedBugs{} previously unknown bugs.
At the time of writing, developers have confirmed \confirmedBugs{} bugs, of which \fixedBugs{} have been fixed.
Among these bug reports, two CVEs have been granted for vulnerabilities in \texttt{sorbet} and \texttt{verible}.

We received positive feedback from LSP server vendors on our bug reports.
Several bugs we reported to \texttt{clangd} developers are buffer-overflow and use-after-free, which can potentially lead to remote code execution when editing malicious source files.
In response to the security advisories we reported (private disclosure), the LLVM team disabled \texttt{clangd} in their \textsc{VSCode} extension for untrusted workspaces~\cite{link:vsc-clangd-812}.
\begin{quote}
	\itshape
	Parsing untrusted code through clang can result in harmful behavior.
	Also it isn't considered as a security-sensitive component, hence its on embedders like \texttt{vscode-clangd} to ensure users are aware of such risks.
\end{quote}

Additionally, during our communication with the \texttt{sorbet} development team, a developer showed interest in integrating \lspfuzz{} into their testing workflow~\cite{link:sorbet-8903}.
\begin{quote}
	\itshape
	I'd love to hear more about what your process looks like for that.
	If there's something repeatable that we could use for our own purposes, I'd love to learn how much effort is involved.
\end{quote}

Our results demonstrate that \lspfuzz{} is not only effective at uncovering previously unknown bugs in production LSP servers, but also valuable for LSP server developers.

\begin{summary}
	\lspfuzz{} can find bugs in real-world LSP servers confirmed by developers.
	We received positive feedback from LSP server developers on our bug reports.
	In addition, they expressed interest in integrating \lspfuzz{} into their testing workflows.
\end{summary}

\subsection{Threats to Validity}

First, we evaluated \lspfuzz{} on four LSP servers, which may not represent the full diversity of LSP servers.
To mitigate this threat, we selected widely used, large-scale, and actively maintained LSP servers.
However, our results may not generalize to all LSP servers, especially those with different implementation strategies.
Evaluating \lspfuzz{} on more LSP servers remains an important direction for future work.

In addition, the results of our experiments may vary between runs due to the inherent randomness of fuzzing.
To mitigate this threat, we repeated each experiment ten times to reduce the impact of randomness.

\section{Related Work}\label{sec:related-work}

As the first LSP server testing technique, our work on \lspfuzz{} draws inspiration from several lines of related work, which we discuss in this section.

\paragraph{Grammar-Based Fuzzing}
The first line of related work is grammar-based fuzzers~\cite{DBLP:conf/pldi/GodefroidKL08,DBLP:conf/ndss/AschermannFHJST19,DBLP:conf/issta/SrivastavaP21,DBLP:conf/esorics/VeggalamRHB16,DBLP:conf/pldi/YangCER11}, which generate structurally sound inputs using formal specifications and significantly improve testing effectiveness.
However, they often fall short because real-world applications like LSP servers require inputs that satisfy complex constraints.
To address this, \textsc{ISLa}~\cite{DBLP:conf/sigsoft/SteinhofelZ22} introduces a specification language for expressing semantic constraints in grammar-based test generation, and \textsc{Fandango}~\cite{DBLP:journals/pacmse/AmayaSZ25} further combines this with search-based techniques for more efficient constraint solving.
Although these techniques can generate inputs with semantic constraints, the lack of a comprehensive LSP constraint set limits their applicability to LSP servers.
In comparison, in \lspfuzz{}, we encode the LSP constraints into the mutation pipeline and perform context-aware mutations, which always lead to test cases conforming to the protocol.

\paragraph{Network Protocol Fuzzing}
The second line of related work focuses on network protocol fuzzing, as LSP is also a protocol despite operating locally.
Several studies have advanced stateful fuzzing by leveraging protocol state identification and efficient state space exploration~\cite{DBLP:conf/icst/PhamBR20,DBLP:journals/tse/MengPBR25,DBLP:journals/ese/Natella22,DBLP:conf/uss/BaBMR22,DBLP:journals/tosem/QinHMZYZ23}.
Other approaches focus on optimizing throughput and discovering memory issues in specialized domains \eg{\textsc{SnapFuzz}~\cite{DBLP:conf/issta/AndronidisC22}, \textsc{IoTFuzzer}~\cite{DBLP:conf/ndss/ChenDZZL0LSYZ18}}.
More recently, \textsc{ChatAFL}~\cite{DBLP:conf/ndss/MengMBR24} leverages LLMs for mutation to escape the coverage plateau.
While these approaches handle protocol state transitions and message sequencing well, they lack awareness of source code, which is essential for LSP servers.
We address this by generating both source code and editor operations, thus covering both protocol and content aspects in our two-stage mutation pipeline.

\paragraph{Compilers and Code Analysis Tools Testing}
The third line of related work concerns testing compilers and code analysis tools, which, like LSP servers, process and analyze source code.
Tools such as \textsc{CSmith}~\cite{DBLP:conf/pldi/YangCER11}, \textsc{LangFuzz}~\cite{DBLP:conf/uss/HollerHZ12}, \textsc{YarpGen}~\cite{DBLP:journals/pacmpl/LivinskiiBR20}, and \textsc{CUDASmith}~\cite{DBLP:conf/compsac/JiangWCTLY020} generate diverse code to test compilers, while \textsc{NSSmith}~\cite{DBLP:conf/asplos/LiuLRTLPZ23} and \textsc{HirGen}~\cite{DBLP:conf/issta/MaS00C23} target deep learning compilers.
Recent work, including \textsc{GrayC}~\cite{DBLP:conf/issta/Even-MendozaSDC23}, \textsc{CovRL-Fuzz}~\cite{DBLP:conf/issta/EomJ024}, and \textsc{Fuzz4All}~\cite{DBLP:conf/icse/XiaPTP024}, applies grey-box fuzzing and LLMs to compiler testing.
While these methods excel at code generation, they overlook the interactive aspect of LSP servers, where bugs may arise from specific combinations of code and editor actions.
Nevertheless, incorporating advanced code generation strategies from these techniques could further enhance \lspfuzz{}.

\paragraph{Multi-Dimensional Fuzzing}
The fourth line of related work examines fuzzing techniques for applications that accept multi-dimensional inputs.
\textsc{Falcon}~\cite{DBLP:conf/issta/YaoHTSWZ21} targets SMT solvers by leveraging the co-existence relation between formula types and configuration options to guide test generation.
However, their approach is too coarse-grained to capture the location-sensitive relationship between code and editor operations in LSP (see \cref{sec:challenge-2}).
\textsc{KextFuzz}~\cite{DBLP:journals/tdsc/YinGXMZZ24} focuses on macOS kernel extensions, employing novel instrumentation, in-kernel interactions, and input format inference to exercise diverse behaviors.
\textsc{DistFuzz}~\cite{DBLP:conf/ndss/ZouBJZZ25} addresses distributed systems by introducing richer event types and pruning communication patterns based on symmetry.
These approaches are either highly specialized for their respective domains or are not intended to address the challenges in LSP server testing (see \cref{sec:formulation}).
Therefore, although LSP servers accept multi-dimensional inputs, these techniques cannot be transferred or adapted for LSP server testing.

\paragraph{LLM-Based Fuzzing}
Large language models are increasingly being adopted in software testing research.
In addition to the previously discussed work~\cite{DBLP:conf/ndss/MengMBR24,DBLP:conf/icse/XiaPTP024,DBLP:conf/issta/EomJ024}, many studies have explored leveraging LLMs for fuzzing.
\textsc{CKGFuzzer}~\cite{new-pubs:gkg-fuzzer} introduces a knowledge graph-driven approach that utilizes LLMs to automate fuzz driver generation and input seed refinement.
Asmita \etal{}~\cite{DBLP:conf/uss/AsmitaOSTFH24} propose two LLM-based techniques to improve fuzz testing of \textsc{BusyBox}.
\textsc{FuzzGPT}~\cite{DBLP:conf/icse/DengXYZY024} employs LLMs to generate edge-case programs for deep learning library testing.
\textsc{PromptFuzz}~\cite{DBLP:conf/ccs/LyuXCC24} uses LLMs for fuzz driver generation.
These studies utilize the empirical knowledge encoded in LLMs to address various challenges in fuzzing.
Given the widespread integration of LLMs in modern code editors, an intriguing direction for future work is to leverage LLMs to generate complex code editing actions, thereby further enhancing LSP server fuzzing.

\section{Conclusion and Future Work}\label{sec:conclusion}

LSP has become the de facto standard for enabling code intelligence features in code editors.
While the reliability of LSP servers is a growing concern, no existing techniques specifically target LSP server testing.
In this paper, we propose \lspfuzz{} to bridge this gap.
\lspfuzz{} outperforms baselines and is effective at detecting bugs and vulnerabilities in real-world LSP servers.
Our bug reports led to prompt actions by LSP server developers to address the issues and received positive feedback.

\lspfuzz{} is not only the first technique of its kind, but also serves as a foundation for further enhancements.
We outline several limitations that present promising future work.

\paragraph{Multi- or Evolving Source File Scenarios}
Our current problem formulation does not target scenarios where LSP servers operate on multiple or evolving source files.
Therefore, \lspfuzz{} cannot detect bugs that are triggered in these scenarios.
Investigating the testing of LSP servers in multi-file workspaces or with dynamically changing source files could uncover bugs related to cross-file dependencies and incremental analysis.
This is challenging due to the wide variety of possible file states and connections, which requires specifically tailored mutation operators, making it an interesting area for future research.

\paragraph{LSP-Specific Test Oracles}
\lspfuzz{} focuses on bugs that lead to crashes or memory corruptions.
However, many functional bugs could be exposed by designing more sophisticated test oracles \eg{go-to-definition requests from multiple references to the same symbol should return the same location}.
Developing and exploring richer test oracles for LSP servers represents an important direction for future work.

\paragraph{Semantic-Aware Source Code Mutation}
Our current approach uses only syntactic mutations, which may not generate sufficiently complex and valid programs as part of the test inputs.
Thus, \lspfuzz{} may miss bugs caused by semantic characteristics \eg{def-use or caller-callee relations}.
Incorporating semantic-aware mutations, inspired by strategies from compiler testing techniques, could help uncover deeper bugs in LSP servers.

\section{Data Availability}
Our artifact is available at the following URL:
\begin{center}
	\artifact{}
\end{center}

\section*{Acknowledgments}

We would like to thank the anonymous reviewers for their insightful comments.
This work is supported by the Hong Kong Research Grants Council General Research Fund (Grant No.
\texttt{16206524}, for authors at HKUST), Natural Sciences and Engineering Research Council of Canada Discovery Grant (Grant No. \texttt{RGCPIN-2022-03744} and \texttt{DGECR-2022-00378}, for authors at McGill), Fonds de recherche du Québec-secteur Nature et technologies (Grant No. \texttt{363482}~\cite{fund:frqnt-new-aca}, for authors at McGill), and the National Natural Science Foundation of China (Grant No. \texttt{62372219}, for authors at SUSTech).

\bibliographystyle{IEEEtran}
\bibliography{bibliography/references,bibliography/links,bibliography/artifact}

\end{document}